\documentclass[%
 aip,
 jcp,%
 amsmath,amssymb,
 reprint,%
final,
]{revtex4-1}
\usepackage{graphicx}
\usepackage{dcolumn}
\usepackage{bm}
\usepackage{color}
\usepackage{xcolor}

\begin{document}

\title{Minima Hopping Guided Path Search: An Efficient Method for Finding Complex Chemical Reaction Pathways} 

\author{Bastian Schaefer}
\email[]{bastian.schaefer@unibas.ch}
\affiliation{Department of Physics, University of Basel, Klingelbergstrasse 82, CH-4056 Basel, Switzerland}
\author{Stephan Mohr}
\affiliation{Department of Physics, University of Basel, Klingelbergstrasse 82, CH-4056 Basel, Switzerland}
\affiliation{Univ. Grenoble Alpes, INAC-SP2M, F-38000 Grenoble, France\\ 
\hspace{1.5mm} CEA, INAC-SP2M, F-38000 Grenoble, France}
\author{Maximilian Amsler}
\affiliation{Department of Physics, University of Basel, Klingelbergstrasse 82, CH-4056 Basel, Switzerland}
\author{Stefan Goedecker}
\email[]{stefan.goedecker@unibas.ch}
\affiliation{Department of Physics, University of Basel, Klingelbergstrasse 82, CH-4056 Basel, Switzerland}

\date{\today}

\begin{abstract}
The Minima Hopping global optimization method uses physically realizable molecular dynamics moves in combination with an energy feedback that guarantees the escape from any potential energy funnel.
For the purpose of finding reactions pathways, we argue that Minima Hopping is particularly suitable as a guide through the potential energy landscape and as a generator for pairs of minima that can be used as input structures for methods capable of finding transition states between two minima.
For Lennard-Jones benchmark systems we compared this Minima Hopping guided path search method to a known approach for the exploration of potential energy landscapes that is based on deterministic mode-following. Although we used a stabilized mode-following technique that reliably allows to follow distinct directions when escaping from a local minimum, we observed that Minima Hopping guided path search is far superior in finding lowest-barrier reaction pathways.
We therefore suggest that Minima Hopping guided path search can be used as a simple and efficient way to identify energetically low-lying chemical reaction pathways.
Finally we applied the Minima Hopping guided path search approach to 75-atom and 102-atom Lennard Jones systems.
For the 75-atom system we found pathways whose highest energies are significantly lower than the highest energy along the previously published lowest-barrier pathway. Furthermore, many of these pathways contain a smaller number of intermediate transition states than the previously publish lowest-barrier pathway.
In case of the 102-atom system Minima Hopping guided path search found a previously unknown and energetically low-lying funnel.
\end{abstract}


\maketitle 

\section{Introduction}\label{Introduction}

The exploration of potential energy landscapes requires two important aspects to be considered.
On the one hand, the geometries of stable ground-states are of large interest.
For this reason powerful global optimization methods such as several genetic algorithms~\cite{Holland1992,Woodley1999,Bazterra2002,Oganov2006,Schoenborn2009}, Basin Hopping~\cite{Wales1997}, the Activation Relaxation Technique~\cite{Barkema1996,Mousseau1998,Malek2000,Machado2011,Mousseau2012} and Minima Hopping (MH)~\cite{Goedecker2004,Roy2008,Schoenborn2009,Amsler2010} have been developed during the last two decades.
On the other hand, processes like  protein folding, catalysis, chemical reactions in solutions and surfaces or the formation of stable phases in solids often force the reacting systems to undergo rarely occurring complex transformations between long-lived states.
Actively stabilizing or destabilizing long-lived states by inhibiting or promoting reaction pathways responsible for certain events allows to synthesize new materials or substances with specifically tailored properties.\cite{Wang2005,Hickenboth2007,Moura2013}
Unfortunately, the sole knowledge of the global minimum and a collection of local minima provided by global optimization methods is not sufficient for being able to influence reaction pathways specifically.
Instead, an accurate knowledge of the atomistic details of reaction pathways is needed.
For this reason, in addition to local minima also transition states and the information which minima are connected by which transition states are of great importance.
As soon as this data is available, various methods like the master equation approach, the discrete path formulation of Discrete Paths Sampling or Kinetic Monte Carlo allow to compute dynamic properties.~\cite{Wales2002, Wales2004, Wales2004-2, Wales2006}
Using graph-theoretic methods it is possible to extract reaction pathways from databases containing the just mentioned data.
Since pathways with energetically high barriers have a vanishingly small contribution to properties like rate constants, it is important not to investigate just any pathways but to sample preferably those that have low overall barriers.

As shown in Ref.~\onlinecite{Doye1999}, such a kind of sampling can in principle be accomplished by mode-following methods coupled to an acceptance-rejection criterion that provides a bias to low-energy configurations.
However, in a study by Doye et al.~\cite{Doye1999-2} a systematic sampling approach was considered not to be able to find even a single pathway connecting both lowest lying minima of the 75-atom Lennard-Jones system within a feasible computation time.
Instead of using a completely unbiased search, they had to use a method which optimizes an initially given input pathway.
The method of constructing an initial pathway which connects two states of interest and subsequently finding lower energy pathways by perturbing the initial path has been used and refined in various ways in later studies conducted by Wales et al..
Apparently, this approach seems to be an efficient procedure for constructing reaction pathways since, in a nutshell, this is the method of choice in the often applied Discrete Path Sampling approach.~\cite{Wales2002,Wales2004-2,Carr2008,Khalili2008}

Conventional methods for computing Hessian eigenvectors (modes) that are based on an iterative minimization of the curvature tend to converge to the lowest Hessian eigenvector, only.
Therefore, deterministic methods using mode-following approaches based on these conventional eigenvector computation methods run into the risk of being non-ergodic, because the number of available escape directions away from a local minimum is very limited.
In section \ref{sec:modefollowing} we show a stabilized mode-following technique that allows to converge reliably to the closest Hessian eigenvector. This somewhat alleviates the problem of converging only to the lowest eigenvector. Therefore it can be used to follow more reliably the full number of $6N-12$ search directions available in a $N$-atomic system (free boundary conditions assumed).

Besides for global optimization, the powerful Activation Relaxation Technique ART nouveau\cite{Malek2000,Machado2011,Mousseau2012} of Mousseau and his coworkers can also be used for computing reaction pathways. In this method, the problem of the restricted number of escape-directions is solved by using random displacements away from the initial local minimum.
ART nouveau has evolved from ART\cite{Barkema1996,Mousseau1998} and has successfully been applied to different systems like for example amorphous\cite{Kallel2010} and crystalline silicon\cite{Levasseur2008a,Ganster2012}, the diffusion of interstitials and vacancies\cite{Levasseur2008b,Marinica2011,Joly2013}, peptides and proteins.\cite{Mousseau2008,Wei2002,Dong2008,StPierre2008,StPierre2012,Dupuis2012}

A further method that has been applied for the calculations of reaction pathways is Transition Path Sampling (TPS) which generalizes importance sampling to trajectory space.~\cite{Dellago1998-1,Dellago1998-2,Bolhuis2002,Dellago2003,Gruenwald2008,Gruenwald2009,Lechner2011}
However, as has been shown by Miller and Predescu, TPS with shooting and shifting moves becomes trapped in high-energy structures of $\mbox{LJ}_{38}$ and thus fails to find the global minimum funnel of this system.
They thus developed a double-ended transition path sampling method, named Sliding and Sampling, which could find pathways between both funnels.~\cite{Miller2007}
However, the main drawbacks of their method are the non-ergodicity of their simulation for $\mbox{LJ}_{38}$ and the high computational cost which is several orders of magnitude higher than that of the above mentioned method by Doye et al.~\cite{Doye1999}

Chemical reaction pathways can be partitioned into a sequence of stationary point crossings. Therefore, many methods that are intended for predicting chemical reaction pathways necessarily must use techniques for converging to stationary points.
However, the main focus of this work is not to compare the efficiency of methods that converge to individual stationary points, but to discuss and benchmark a new scheme for generating sequences of stationary points from which low-barrier pathways leading over many barriers can be extracted.
To do so, we re-examine a systematic potential energy landscape exploration method that has been outlined in Ref.~\onlinecite{Doye1999}. In contrast to Ref.~\onlinecite{Doye1999}, we use a stabilized mode-following method which is introduced in section \ref{sec:modefollowing}. Although, this stabilized mode-following method alleviates the problem of preferentially escaping a minimum along the lowest Hessian eigenvector only, we come to similar results as previous investigations:~\cite{Doye1999-2,Malek2000} We conclude that in general this systematic potential energy landscape exploration approach is not optimal and occasionally fails to find lowest-barrier pathways for even moderately sized systems like $\mbox{LJ}_{38}$.

By virtue of the explosion condition~\cite{Goedecker2004,Goedecker2011} MH is guaranteed to escape from any potential energy funnel and due to the molecular dynamics (MD) based moves the minima along the MH trajectory are separated by low energy barriers. Furthermore, the consecutive minima are structurally not too different from each other, because the MD moves consist of a few steps, only. These properties make MH particularly suitable to serve as a guide for searching low-energy reaction pathways. These pathways can connect parts of the potential energy landscape that are far away from each other and that are possibly separated by high energy barriers.
Combining MH and a suitable method for finding transition states between two input geometries leads to the novel Minima guided path search (MHGPS) approach presented in section \ref{sec:MHGPS}.
Using MHGPS we mapped out the energy landscape of $\mbox{LJ}_{75}$ and $\mbox{LJ}_{102}$.
Despite numerous published investigations of the Lennard-Jones clusters, we were able to detect many pathways that are significantly lower in energy and shorter with respect to the integrated path length and number of intermediate transition states than previously known pathways for $\mbox{LJ}_{75}$.~\cite{Doye1999-2}
For $\mbox{LJ}_{102}$ we found a third, previously unknown and energetically low-lying funnel at the bottom of which a new structural motif is located.
The pathways found between both lowest minima of $\mbox{LJ}_{102}$ are also significantly shorter in terms of the number of intermediate transition states and in terms of the integrated path length when compared to previously presented pathways.~\cite{Doye2006}

\section{Methods}
\subsection{Lennard-Jones Potential}\label{sec:LJ}
All  interactions in this study were modeled by the Lennard-Jones (LJ) potential~\cite{Lennard-Jones1924,Lennard-Jones1925}
$$E = 4 \epsilon \sum_{i<j}\left\{\left(\frac{\sigma}{r_{ij}}\right)^{12}-\left(\frac{\sigma}{r_{ij}}\right)^6\right\},$$
where $\epsilon$ defines the pair-well depth and $2^{1/6}\sigma$ is the pair-well equilibrium distance.
All energies and distances are reported in units of $\epsilon$ and $\sigma$, respectively.
\subsection{Transition states, their connectivity and stationary point databases}\label{sec:transitionstates}
We follow the usual definition of a transition state being a first order saddle point of the energy function.~\cite{Wales2004}
Steepest descent paths connect transition states to two stationary points.
In most cases these stationary points are local minima.
We adapt the terminology of Wales~\cite{Wales2002,Wales2004,Wales2004-2} and denote sequences of minima and transition states connected by steepest descent paths as `discrete paths'.
A collection of local minima, transition states and the information which transition states connect which minima is called a `stationary point database'.~\cite{Wales2002,Wales2004,Wales2004-2} 

Building stationary point databases requires the identification or distinction of atomic configurations with or from each other.
For this purpose we utilized the recently developed fingerprints which are based on the eigenvalues of a s-orbital overlap matrix.~\cite{Sadeghi2013}
For the calculation of the fingerprints, we used $2^{1/6}\frac{\sigma}{2}$ as the covalent radius of the LJ atoms.
We considered two conformers to be identical if their energy difference was smaller than $10^{-5} \epsilon$ and their fingerprint distance less than $2\times10^{-4}$.

Extracting from a stationary point database all lowest-barrier paths with the least number of intermediate transition states between two given minima poses a problem that is closely related to the so called shortest-widest~\cite{Ma1997} path problem.
This can be solved by applying Dijkstra's algorithm~\cite{Dijkstra1959} twice.~\cite{Ma1997}
In the first step Dijkstra's algorithm searches for all paths that connect both minima with the lowest possible energy barrier $E_{\mbox{barr;lowest}}$.
The stationary point database then is truncated by removing all transition states with energies higher than $E_{\mbox{barr;lowest}}$.
Next, Dijkstra's algorithm passes through the truncated database and searches for the path with the smallest possible number of intermediate transition states.

To determine the connectivity in all sampling approaches presented below, we stepped away from a transition state by adding to and subtracting from the transition state one-100th of the normalized Hessian eigenvector that corresponds to the negative curvature.
Using Euler's method with a maximum step size of $10^{-2} \sigma$, approximate steepest descent paths were computed until the Euler integrator entered the quadratic region surrounding a minimum.
In this Euler integration scheme steps were rejected and the step size was decreased if either the angle between the gradients of two successive steps was larger than 60 degree or if the energy increased.
Inside the quadratic region the Euler method was replaced by the fast inertial relaxation engine (FIRE)~\cite{Bitzek2006} in order to speed up the geometry optimization.
For the FIRE integrator itself it is not of any relevance whether it operates inside the quadratic region or not.
However, compared to non-quadratic regions it seems less likely that inside the quadratic region the FIRE method will converge to a different minimum than Euler's method.
Because dynamic properties computed from stationary point databases are unlikely to depend strongly on whether the connectivity of the potential energy landscape is established by using approximate steepest descent paths or paths from advanced minimization algorithms~\cite{Wales2004, Wales2006} like for example FIRE or the Broyden–Fletcher–Goldfarb–Shanno (BFGS) algorithm~\cite{Broyden1970,Fletcher1970,Goldfarb1970,Shanno1970,press_numerical_1992}, the time used for relaxations to local minima could have shortened significantly when omitting the Euler integration and using advanced minimization algorithms throughout.
However, because we introduce a new reaction pathway search method, we decided to use the conservative Euler integration approach in order to sample connectivity information that is in accordance with the connectivity defined by the widely accepted intrinsic reaction coordinate.~\cite{Fukui1970}
Although we do not report any results based on FIRE-only minimization, we compared the differences of pathways obtained from FIRE-only and Euler integration plus FIRE optimization.
We only observed changes in the number of intermediate transition states.
In all cases the energetically lowest transition state between two states found by FIRE-only runs was identical to the lowest transition state found by connections established by approximate steepest descent paths.

In addition to the conservative combination of Euler's method and FIRE, all new pathways explicitly reported in this study (Figures~\ref{fig:pathslj75} and \ref{fig:pathslj102}) were double-checked in a post-processing step.
In order to obtain quasi-exact intrinsic reaction pathways, steepest descent paths were recomputed using only Euler's method with a maximum displacement of $10^{-6} \sigma$ in each integration step.
Before this steepest descent relaxation the structures had been pushed away from the transition state one-10,000th of the normalized eigenvector belonging to the negative Hessian eigenvalue.

It has to be emphasized that, similar to all commonly used global optimization algorithms, the methods presented in this work do not rigorously guarantee that an optimal solution has been found.
That is, all presented structures and lowest-barrier pathways should be denoted as `putative lowest structures' or `putative lowest-barrier pathways'.
However, for convenience we sometimes omit the word `putative'.

\subsection{Disconnectivity graphs}\label{sec:disconnectivitygraphs}
Disconnectivity graphs introduced by Becker and Karplus~\cite{Becker1997} and frequently used and illustrated by Wales et al.~\cite{Wales1998,Wales2004,Wales2006,disconnectiondps} can be used to visualize stationary point databases of multidimensional potential energy landscapes.
They therefore allow to obtain a rough, intuitive insight into dynamic properties.
In this section we briefly recapitulate the theory of disconnectivity graphs.

Disconnectivity graphs illustrate which minima are convertible into each other by following reaction pathways without ever exceeding a given threshold energy.
Such mutually accessible regions are called 'superbasins'.~\cite{Wales2006}
The number of superbasins depends on the threshold energy.
The vertical axis of a disconnectivity graph is partitioned into a predefined and freely chosen number of energy thresholds.
At each threshold energy the superbasins are represented by nodes on the graph and are arranged along the horizontal axis.
At threshold energies at and above which superbasins are mutually accessible, the corresponding nodes below this threshold energy are connected by lines.
Finally all the single minima at the bottoms of the superbasins are represented separately by drawing lines down to the energy of each minimum.
The horizontal position of the nodes and minima is arbitrary.
Typically there are too many minima to visualize, hence only the lowest $n$ minima are usually plotted.
Nevertheless, all minima and transition states contained in the underlying stationary point database contribute to the superbasin and barrier analysis.

The number and positions of the chosen threshold energies can heavily influence the appearance of a disconnectivity graph and hence these parameters have to be well chosen in order to obtain a suitable trade-off between a detailed and coarse grained visualization of the topological information contained in the underlying stationary point database.~\cite{Wales2006}

The plots of all disconnectivity graphs in this work were generated using the disconnectionDPS~\cite{disconnectiondps} software.

\subsection{A stabilized mode-following method}\label{sec:modefollowing}
\begin{figure}
 \includegraphics{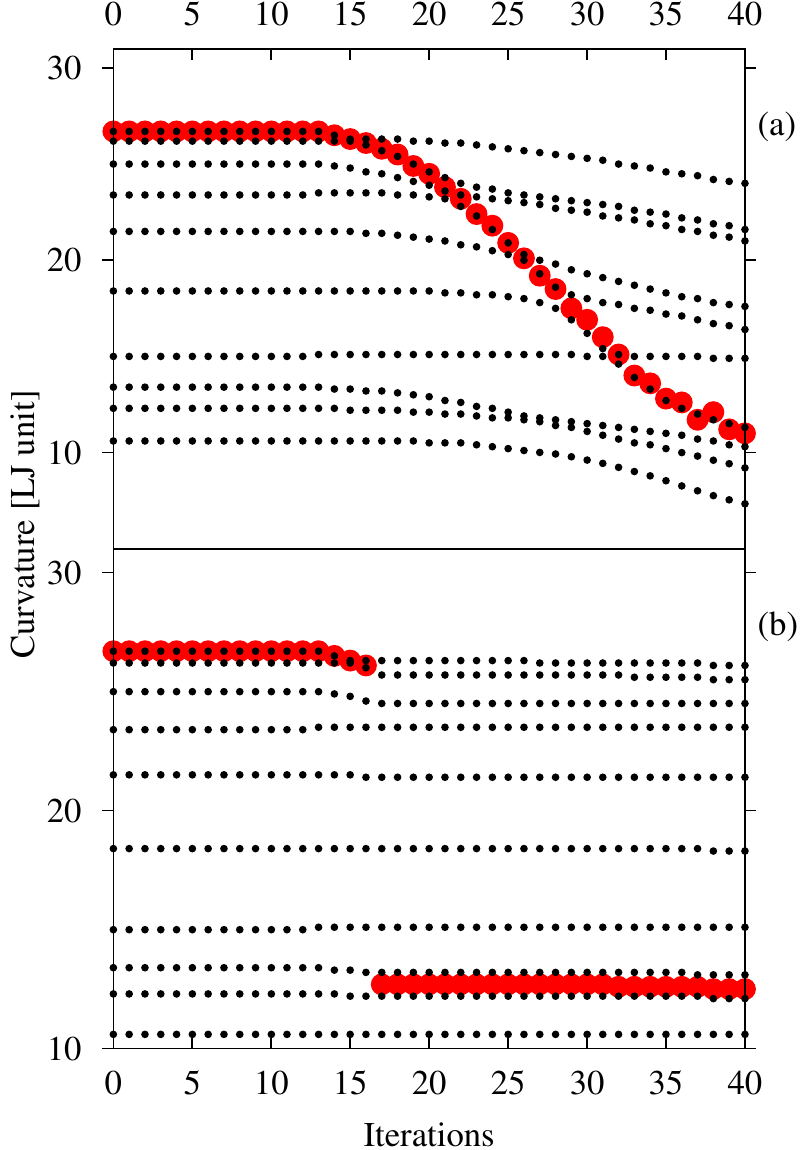}
 \caption{A visualization of how using DIIS for the dimer rotations helps to stay on a given mode. Panel (a) visualizes data obtained by using DIIS, panel (b) shows data obtained by using steepest descent.
The small black dots are the ten lowest eigenvalues of the Hessian at each step of a trajectory starting at a local minimum, whereas the large red dots are the curvature along the search direction. 
The DIIS procedure in panel (a) stays in general on the mode that has the largest overlap with the dimer direction, and thus stays on the initial mode for quite a long time.
In contrast to this, the steepest descent procedure in panel (b) becomes unstable as soon as the 9th and 10th mode cross and switches to a low curvature mode, as a consequence.
}
 \label{fig:new_mode_following}
\end{figure}

As the name suggests, the basic idea of mode following methods is to find the path from a minimum to a first-order saddle point by following an eigenmode of the Hessian.\cite{Wales1994,Wales1996,Munro1999,Wales2004}
In practice the determination of the eigenmodes via a diagonalization is too costly and one therefore has to resort to iterative methods, 
meaning that the mode to be followed is found by a minimization problem.

In our approach the Hessian eigenmodes are found using a version of the dimer method~\cite{Henkelman1999}. The dimer consists of two images $\mathbf{R}_1$ and $\mathbf{R}_2$ in the $3N$-dimensional search space, 
separated by a short distance $2\epsilon$:
\begin{align}
 \mathbf{R}_1&=\mathbf{R}_0+\epsilon\hat{\mathbf{N}}\\
 \mathbf{R}_2&=\mathbf{R}_0-\epsilon\hat{\mathbf{N}}
\end{align}
where $\hat{\mathbf{N}}$ is the normalized dimer direction and $\mathbf{R}_0$ is the dimer midpoint.
The dimer method  first rotates the dimer in order to align it with a Hessian eigenmode 
and then translates it along this mode.
This procedure is repeated until the transition state is reached.
As explained below, conventional methods that compute Hessian eigenvectors by an iterative minimization tend to converge preferentially to the eigenvector corresponding to the lowest Hessian eigenvalue.
However, in order to escape from a minimum to many different transition states, it is desirable to follow as many different escape directions as possible in the beginning of the mode-following procedure.
The ability to converge reliably to the closest Hessian eigenvector, and thus being able to systematically follow many different directions is the original contribution of the method presented in this section.

\subsubsection{Rotating the dimer}
The essential point of the dimer method is to find an efficient prescription for the rotational part.
The quantity that has to be minimized is the curvature along the dimer direction, 
$C_{\mathbf{R}_0}(\hat{\mathbf{N}})=\hat{\mathbf{N}}^T\mathbf{H}_{\mathbf{R}_0}\hat{\mathbf{N}}$, 
where $\mathbf{H}_{\mathbf{R}_0}$ is the Hessian evaluated at the dimer midpoint.
Since the computation of the exact Hessian is in general too costly, 
the curvature is approximated using finite differences computed from the forces that act on the two images of the dimer~\cite{Henkelman1999}:
\begin{equation}
 C_{\mathbf{R}_0} = \frac{\left(\mathbf{F}_2-\mathbf{F}_1\right)\cdot\hat{\mathbf{N}}}{2\epsilon}.
\end{equation}

There are ways to locally approximate the curvature by a short Fourier series and then to directly minimize this expression~\cite{Heyden2005}. 
However we chose a more straightforward approach by working directly with the torsional force~\cite{Henkelman1999}
\begin{equation}
 \mathbf{F}^\perp = (\mathbf{F}_1-\mathbf{F}_2) - \left(\left(\mathbf{F}_1-\mathbf{F}_2\right)\cdot\hat{\mathbf{N}}\right)\hat{\mathbf{N}}.
\end{equation}
Image 1 of the dimer is now iteratively moved according to this force until the latter falls below a given threshold; 
at each step the position of the image has to be adjusted to keep the dimer separation constant. 
In order to reduce the number of force evaluations, 
the force acting on image 2 is approximated by using the force acting on the dimer midpoint~\cite{Olsen2004,Heyden2005}, $\mathbf{F}_0$, i.e.\ $\mathbf{F}_2=2\mathbf{F}_0-\mathbf{F}_1$, 
in this way leading to
\begin{equation}
 \mathbf{F}^\perp = 2(\mathbf{F}_1-\mathbf{F}_0) - 2\left(\left(\mathbf{F}_1-\mathbf{F}_0\right)\cdot\hat{\mathbf{N}}\right)\hat{\mathbf{N}}.
\end{equation}
Since the value of $\mathbf{F}_0$ does not change during the rotation, only one force evaluation per iteration is required.

For the current purpose it is crucial to have the ability to find systematically many different transition states leading out of a given minimum.
This means that one has to be able to follow many different modes.
For the mentioned dimer method -- and as well for other related mode following methods~\cite{Wales1994,Wales1996,Munro1999,Wales2004} -- this is not the case. The reason for this is very simple. 
As is shown in Appendix \ref{app:Stability_of_the_modes}, 
only the lowest mode is a stable one, meaning that the curvature has a local minimum there. All other modes represent saddle points (except for the highest mode which is a maximum). 
This implies that, as soon as the search mode deviates from an exact eigenmode of the Hessian -- which will inevitably happen during a mode following process due to the finite step size for the translation --
there is a strong tendency that the re-determination of the exact eigenmodes will lead to the lowest one, even though one might initially have been aligned along another one. 
In other terms, it is very likely that searches started along different modes of a given minimum will lead to the same saddle point and thus the efficiency of exploring the potential energy landscape and finding lowest-barrier reaction pathways is degraded.

This problem can be circumvented by a very simple modification, namely by using direct inversion of the iterative subspace (DIIS)~\cite{Pulay1980} to perform the rotation of the dimer. 
Since DIIS has the tendency to find the closest stationary point\cite{Machado2011}, 
the iterative procedure to come back to the exact eigenmode will not lead to the lowest mode, 
but rather to the one which has the largest overlap with the previous one. 
In this way the dimer method is stabilized and it is possible to systematically follow different modes out of a given minimum.
In order to avoid any instabilities related to the DIIS procedure, it is required that the starting point does not lie too far away from the exact eigenmode.
This is achieved by keeping the step size for the translation reasonably small.
A comparison of the stabilized mode following technique using DIIS and a standard approach using steepest descent for the rotations is shown in Fig.~\ref{fig:new_mode_following}.

It has to be emphasized that ART nouveau~\cite{Mousseau2012} also use DIIS.
However, in contrast the method described here, ART nouveau uses DIIS in order to move on the potential energy landscape towards transition states, whereas in this section DIIS is used to rotate the dimer.

As one is searching first-order saddle points, it is necessary that one finally ends up on the lowest mode, no matter which mode one has started with. 
It turns out that the order of the mode usually decreases as one moves away from the minimum, 
but in order to safely reach a saddle point it is still necessary to abandon the initial mode at some point 
and to follow the lowest mode instead -- a simple criterion to do so is when the second derivative of the energy with respect to the number of iterations becomes negative.
In our implementation the lowest mode was determined by using the Lanczos method~\cite{lanczos1988applied}, as presented in Ref.~\onlinecite{Olsen2004}.

\subsubsection{Translating the dimer}
In contrast to the rotation of the dimer, the translation is rather straightforward, following the approach outlined in Ref.~\onlinecite{Henkelman1999}.
If the saddle point search was started from a local minimum, then there are two cases to distinguish. 
First the dimer has to be brought out of the convex region around the minimum. To this end it is moved upwards along the dimer direction using the most simple prescription, i.e. $\mathbf{R}_0^{(i+1)}=\mathbf{R}_0^{(i)}+\alpha\mathbf{F}_{eff}^{(i)}$ with $\mathbf{F}_{eff}^{(i)}=-(\mathbf{F}^{(i)}_{0}\cdot\hat{\mathbf{N}}^{(i)})\hat{\mathbf{N}}^{(i)}$, $\alpha>0$.
This is the method of choice until the curvature along the dimer axis becomes negative. As soon as this happens, the effective force is altered to $\mathbf{F}_{eff}^{(i)}=\mathbf{F}^{(i)}_{0}-\lambda(\mathbf{F}^{(i)}_{0}\cdot\hat{\mathbf{N}}^{(i)})\hat{\mathbf{N}}^{(i)}$, where $\lambda$ was typically set to 10.
In this way the dimer will be guided towards the saddle point. However the procedure can be become inefficient as soon as the dimer is close to the stationary point.
In this case it is advisable to switch to a convergence accelerator; in our case we were using DIIS, an approach which is also employed in ART nouveau~\cite{Mousseau2012}.

\subsection{Generating stationary point databases using the mode-following approach}
In Ref.~\onlinecite{Doye1999} Doye et. al. presented an algorithm that allows mode-following techniques to be used for the exploration of the potential energy landscape.
In order to map out the potential energy landscape, we used this algorithm in conjunction with our stabilized mode-following method.
Based on the method used for the transition state search, we henceforth will denote the potential energy landscape exploration method of  Ref.~\onlinecite{Doye1999} as the eigenvector following exploration (EFE) method.
For the convenience of the reader we hereafter shortly recapitulate the EFE method.
Concisely spoken, the walker of the EFE method starts at a local minimum and follows the lowest Hessian eigenvector until a transition state is found.
If this transition state leads to a minimum with an energy that is less than or equal to the energy of the current minimum, the new minimum is accepted and a new transition state search is initiated from this minimum.
If a transition state leads to a minimum that is higher in energy, or if the transition state is not connected to the current minimum, the move is discarded and a further transition state search is begun at the current minimum, either by following the negative direction of the just followed mode, or if this already has been done, by following the direction of the eigenvector belonging to the next higher Hessian eigenvalue.
For each minimum, only a maximum number of transition state searches is performed (less or equal than $6N-12$, where $N$ is the number of atoms).
If this number is exceeded, no new transition state searches are initiated from this minimum and the algorithm jumps to the minimum that is next higher in energy and for which the maximum number of transition state searches have not been accomplished yet.

\subsection{Bar-Saddle}\label{sec:barsaddle}
 \begin{figure}
 \includegraphics{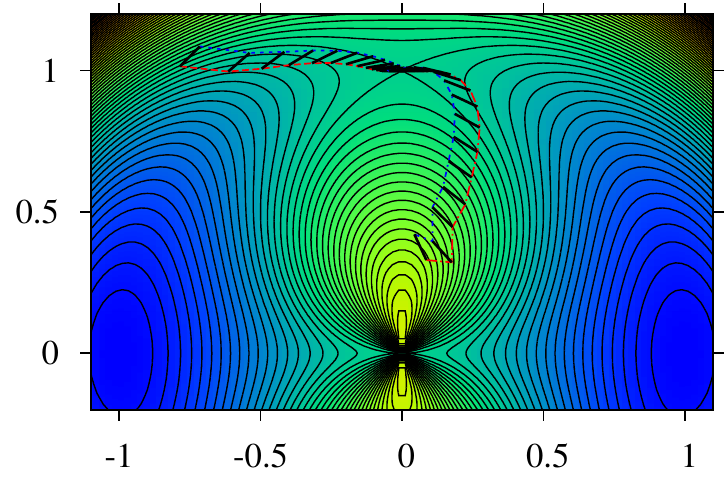}%
 \caption{\label{fig:barsaddle} Trajectory of the bar starting from two initial positions on a model energy landscape  $f(x,y)=(1-(x^2+y^2))^2+(y^2)/(x^2+y^2)$. Two minima are located at $(\pm1,0)$, and the saddle point is located at $(0,1)$.}
 \end{figure}

In the presence of friction, a ball released from a high altitude mountainside would roll downhill and lead to a close-by local minimum.
The Bar-Saddle method presented here uses the idea that, in contrast to the rolling ball, a solid, horizontal bar would roll to the closest saddle point if its point of contact with the surface is kept at its center.
In our implementation a bar is represented by two endpoints $A$ and $B$ at the coordinates $\mathbf{R}_{A}$ and $\mathbf{R}_{B}$ in the high-dimensional configuration space.
The length of a bar is evaluated as $h=|\textbf{R}_{\overline{AB}}|=|\mathbf{R}_B-\mathbf{R}_A|$.
Although the Bar-Saddle formalism derived below is formally closely related to the dimer method~\cite{Henkelman1999},  it follows a different usage paradigm.
The Bar-Saddle formalism can be used to find transition states connecting two given minima.
To do so, it starts from a configuration that is geometrically in between the two input minima and high in energy.
In principle the highest energy configuration along the linear interpolation path between two minima can be used.
However in order to avoid colliding atoms we prefer the freezing string method in Cartesian coordinates for identifying a high energy geometry.~\cite{Behn2011}
In all computations we used a new-node interpolation distance corresponding to 1/10th of the Euclidean distance of the given two minima.
Perpendicular relaxations were stopped as soon as the perpendicular force fell below 5$\frac{\epsilon}{\sigma}$ or as soon as the iteration counter for the perpendicular relaxations was equal to four.
Configurations in between the nodes generated by the freezing string method were interpolated using a cubic spline interpolation. 
A maximum energy configuration along this interpolated path was searched using Brent's method\cite{Brent1972} in between each pair of nodes and then selecting the energetically highest configuration that was found.
Section \ref{sec:MHGPS} describes how to obtain two suitable local minima which serve as input for the freezing string method.

Having identified a suitable starting configuration from which the bar can roll down, the bar is moved iteratively such that the maximum energy along the direction of the bar is at its center (corresponding to the point of contact) and such that the energy at its center is minimized along all directions perpendicular to the bar.
In each iteration, the energies and the forces are evaluated at the bar ends.
The forces are then decomposed into a component parallel to the bar $\mathbf{F}_{\text{i}}^{\parallel}=\left(\mathbf{F}_{\text{i}}\cdot \hat{\mathbf{h}}\right)\hat{\mathbf{h}}$ and a component perpendicular to the bar $\mathbf{F}_{\text{i}}^{\perp}=\mathbf{F}_{\text{i}}-\mathbf{F}_{\text{i}}^{\parallel}$, where $i=A,B$ and  $\hat{\mathbf{h}}=\hat{\mathbf{R}}_{\overline{AB}}$ is the unit vector along the bar.

For the translation of the bar its energy and force along the bar is defined by a cubic interpolation at the center of the bar, such that
\begin{align}
E_{h/2}&=\frac{1}{8} (4 E_A + 4 E_B + (f_B - f_A) h),
\end{align}
and
\begin{align}
\mathbf{F}_{h/2}^{\parallel}&=\frac{6 E_A - 6 E_B - (f_A + f_B) h}{4 h} \hat{\mathbf{h}},
\end{align}
where $f_i=\mathbf{F}_{\text{i}}\cdot \hat{\mathbf{h}}$.

The perpendicular force is evaluated by $\mathbf{F}_{\text{h/2}}^{\perp}=\frac{1}{2}(\mathbf{F}_{A}^{\perp}+\mathbf{F}_{B}^{\perp})$, such that the total translational forces on the bar ends result to $\mathbf{F}_{A}^{\text{Trans}}=\mathbf{F}_{B}^{\text{Trans}}=-\gamma\mathbf{F}_{h/2}^{\parallel}+\mathbf{F}_{\text{h/2}}^{\perp}$, where $\gamma > 0$. In our implementation we chose $\gamma = 2$.

In addition, a rotational force is applied to the bar in order to approximately align it along the lowest curvature direction.
This additional force is given by $\mathbf{F}_{A}^{\text{Rot}}=\frac{1}{2}(\mathbf{F}_{A}^{\perp}-\mathbf{F}_{B}^{\perp})$ and $\mathbf{F}_{B}^{\text{Rot}}=\frac{1}{2}(\mathbf{F}_{B}^{\perp}-\mathbf{F}_{A}^{\perp})$.

Finally, following a steepest descent approach, the bar ends are moved along the effective forces $\mathbf{F}_{i}^{\text{Eff}}=\alpha\mathbf{F}_{i}^{\text{Trans}}+\beta\mathbf{F}_{i}^{\text{Rot}}$, where $\alpha>0$ and $\beta>0$ define the translational and rotational step sizes.
After each step, the bar length is rescaled such that the new bar length remains the same in each iteration $|\mathbf{R}_B^{\text{New}}-\mathbf{R}_A^{\text{New}}|\stackrel{!}{=}h$.

In comparison to Bar-Saddle, the dimer method estimates both the parallel and perpendicular components of the translational force by the arithmetic mean of the forces at the dimer endpoints. The force responsible for the rotation acts only on one endpoint in case of the dimer method and the rotation is implemented by using the parametrization of a circle in a 2-dimensional plane and rotating the dimer in a single step by an angle estimated using a modified one-dimensional Newton method.~\cite{Henkelman1999}

Fig.~\ref{fig:barsaddle} shows the trajectories of the Bar-Saddle method on a model energy landscape. Note that, although the method works most efficiently if the initial point is energetically higher than the saddle point, it will still converge when the search is started close to a local minimum.

The efficiency of the method can be improved by applying an energy or gradient feedback to the step sizes $\alpha$ and $\beta$.
In practice we used a hybrid method where the first few iterations were obtained from steepest descent with gradient feedback, followed by a BFGS minimization~\cite{Broyden1970,Fletcher1970,Goldfarb1970,Shanno1970,press_numerical_1992} with respect to the translational force $\mathbf{F}_{i}^{\text{Trans}}$ only and applying the rotational forces separately in each iteration.

In our implementation we considered a Bar-Saddle computation as converged if the force norm at the center of the bar fell below $10^{-5}\frac{\epsilon}{\sigma}$ and the curvature in bar direction was negative.
Typically, only on the order of $0.1\%$ of all saddle computations used for the simulations reported in this study could not meet these convergence criteria within 15,000 iterations.

\subsection{Generating stationary point databases using the Minima Hopping guided path search approach}\label{sec:MHGPS}
 \begin{figure*}
 \includegraphics{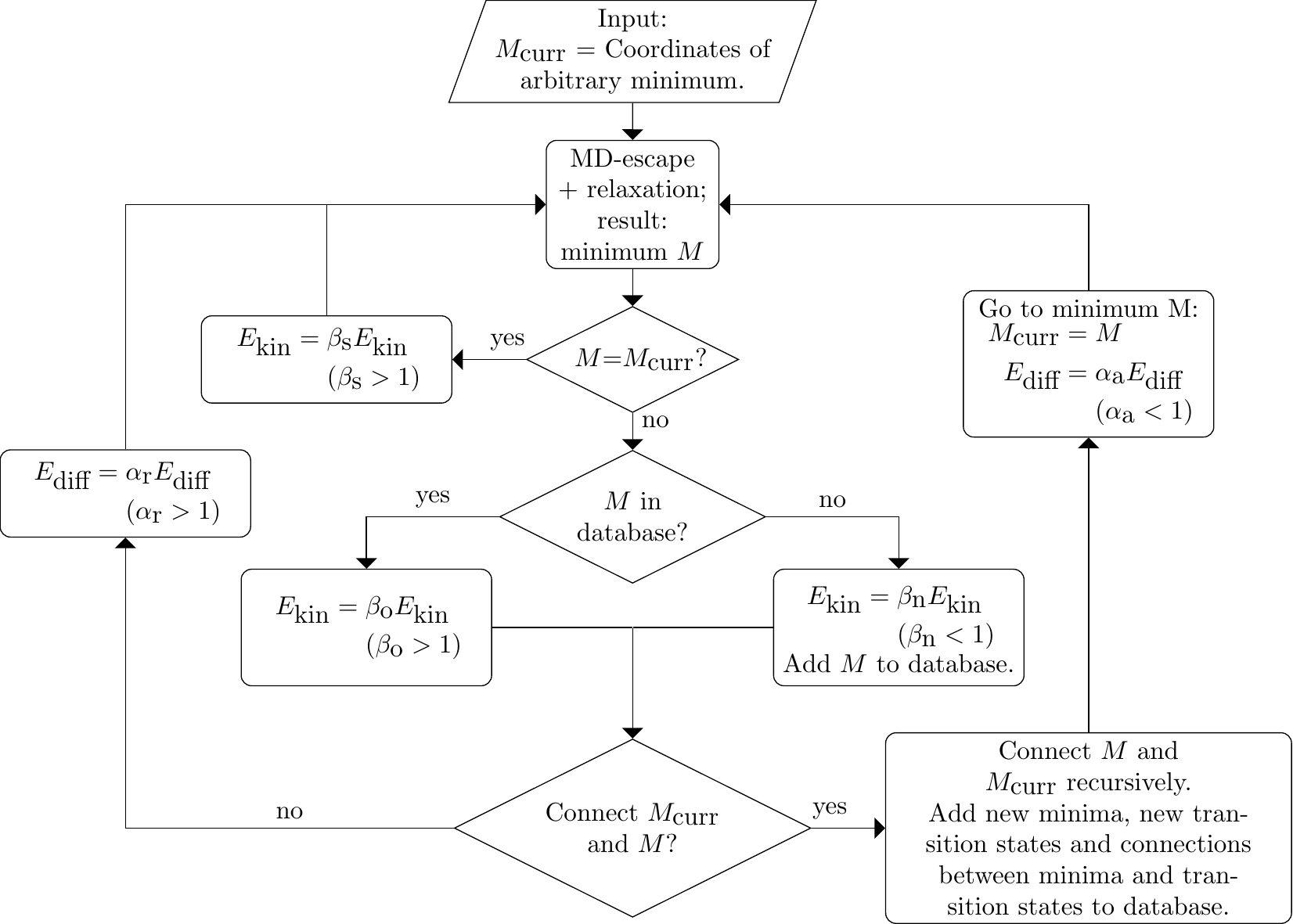}%
 \caption{\label{fig:flowchartMDexplorer}Flowchart describing the new MHGPS scheme. An explanation of the single steps is given in the text of section \ref{sec:MHGPS}.}
 \end{figure*}
Searching for reaction pathways and the exploration of the connectivity of energy landscapes requires an algorithm that moves efficiently inside one funnel and between several funnels. An algorithm that has proven its efficiency in exploring the low energy regions of potential energy landscapes is MH.~\cite{Goedecker2004,Goedecker2005,Hellmann2007,Roy2008,Bao2009,Schoenborn2009}
The success of MH relies in a large part on the MD-type moves and on an energy feedback which satisfies the explosion condition~\cite{Goedecker2004,Goedecker2011}.
The MD moves assure that only physically realizable structures are explored and by means of energy conservation only low-energy barriers are surmounted in unexplored regions of the potential energy landscape.
In well explored regions the explosion condition rigorously guarantees an exponential increase of the kinetic energy.
Therefore, in contrast to most other landscape exploration methods, MH is able to escape automatically from any funnel, irrespective of its depth.
In general, the MD trajectories of MH are short and therefore one can expect consecutive minima along the MH trajectory to be structurally similar to and well aligned with each other.
This alleviates the process of finding intermediate transition states without the need of an explicit and computationally expensive optimization of the geometric and permutational structural alignment\cite{Sadeghi2013}.
In conclusion, MH explores potential energy landscapes efficiently, without the risk of getting trapped and at the same time generates consecutive minima that are particularly suitable for the input of methods that are intended to find transition states located between the input minima.
It seems therefore natural, to combine the  capabilities of MH with a method that connects two given minima by a series of transition states to a Minima Hopping guided path search (MHGPS) technique.

Fig.~\ref{fig:flowchartMDexplorer} shows a flow chart of our new  MHGPS approach.
Just like MH, MHGPS begins at a local minimum and tries to escape from its catchment basin by following a short, random and soft mode biased MD trajectory at the end of which a local geometry optimization is performed.
The softening procedure has been described previously.~\cite{Schoenborn2009}
The escape trials are repeated until MHGPS successfully escapes from the catchment basin of the current minimum.
In order to avoid of getting trapped in the current catchment basin, the kinetic energy is  increased by a factor $\beta_{\mbox{s}}$ after each failed escape trial.
When MHGPS successfully escapes to a different minimum it either decreases the kinetic energy by a factor $\beta_{\mbox{n}}$ or increases it by a factor $\beta_{\mbox{o}}$, depending on whether the new minimum has been visited before or not.
This introduces a feedback which promotes cooling down in unexplored regions and heating up in well explored regions of the potential energy landscape and thus ensures that the algorithm quickly samples the bottom of a funnel and at the same time does not get trapped.

Based on a Metropolis-like\cite{Metropolis1953} criterion MHGPS decides whether it should connect the current minimum  $M_{\mbox{curr}}$ and the new minimum $M$ by a discrete path.
If the energy of the new minimum $E$ is lower than the energy $E_{\mbox{curr}}$ of the current minimum, a connection attempt is always made.
If its energy is higher than the energy of the current minimum, an attempt is made with a probability of
\begin{equation}
\exp\left(-\frac{E-E_{\mbox{curr}}}{E_{\mbox{diff}}}\right).
\end{equation}
The parameter $E_{\mbox{diff}}$ resembles the energy $k_{\mbox{B}}T$ of an ordinary Metropolis simulation. However, in contrast to an ordinary Metropolis simulation, $E_{\mbox{diff}}$ constantly gets adjusted.  If the decision is made to connect $M_{\mbox{curr}}$ and $M$, $E_{\mbox{diff}}$  is decreased by a factor $\alpha_{\mbox{a}}<1$, otherwise it is increased by a factor $\alpha_{\mbox{r}}>1$.

The connections are made by recursively applying Bar-Saddle and following approximate steepest descent paths from emerging intermediate transition states.
Establishing the connection between the two Bar-Saddle input minima $M_{\mbox{curr}}$ and $M$ in a recursive or iterative fashion is essential, because there is no guarantee that the two minima $M_{\mbox{curr}}$ and $M$ can be connected with each other by exactly one transition state.
Hence, during a connection intermediate transition states can appear which might not be connected to one or to both of the two input minima.
In such a case the minima to which the intermediate transition states are connected also have to be connected to the corresponding Bar-Saddle input minima in order to obtain a discrete path that properly connects $M_{\mbox{curr}}$ and $M$.

After connecting $M_{\mbox{curr}}$ and $M$ by a discrete path, the new minimum becomes the current one and the algorithm starts a new MD trajectory at this minimum.
The whole procedure is stopped as soon as a given number of distinct minima are identified.
In all simulations presented in this study  the standard minima hopping parameters ($\beta_{\mbox{s}}=\beta_{\mbox{o}}=1/\beta_{\mbox{n}}=\alpha_{\mbox{r}}=1/\alpha_{\mbox{a}}=1.05$) were used.~\cite{Schoenborn2009,Goedecker2011}

MHGPS is not limited to using Bar-Saddle for connecting minima.
In principle any saddle search method that can find transition states between two given minima, like for example the Nudged Elastic Band method~\cite{Mills1994,Mills1995,Henkelman2000a,Henkelman2000b} or the Splined Saddle method~\cite{Granot2008,Ghasemi2011} can be used.
We decided to use the Bar-Saddle method, because it was the most reliable implementation available to us.

It must be emphasized that, when used alone, methods like the Nudged Elastic Band method or the Splined Saddle method are not suitable for finding lowest-barrier pathways or pathways between structurally very different configurations.
These methods often fail to find a connection between distant minima and, in the best case, can only find some pathway, but not a path having a low overall-barrier.\cite{Koslover2007}

\section{Benchmarks and comparisons}\label{sec:analysis}
\begin{table}
\caption{Results of performance test for $\mbox{LJ}_{38}$.
Averages for $\left<n_{\mbox{ts,diff}}\right>$  and $\left<n_{\mbox{ts}}\right>$ are taken over $1000-n_{\mbox{f}}$ independent and successful runs.}
\begin{ruledtabular}
\begin{tabular}{c c r r r r}
Method  & $n_{\mbox{ev}}$\footnotemark[1] &$\left<n_{\mbox{ts,diff}}\right>$\footnotemark[2] & $\left<n_{\mbox{ts}}\right>$\footnotemark[3] & $n_{\mbox{E}}$\footnotemark[4] & $n_{\mbox{f}}$\footnotemark[5]\\
\hline
MHGPS & n/a & 9267 & 14580 & 3464 & 0\\
EFE & 10 & 64611   & 168688 & 3384 & 24\\
EFE & 25 & 72977   & 192097 & 3508 & 8 \\
EFE & 40 & 91313   & 268422 & 3492 & 1\\
\end{tabular}
\end{ruledtabular}
\footnotetext[1]{Number of lowest eigenvectors along which transition states were searched in positive and negative direction}
\footnotetext[2]{Average number of distinct transition states needed to be found before identifying a lowest-barrier pathway.}
\footnotetext[3]{Average number of transition states computations needed before identifying a lowest-barrier pathway.}
\footnotetext[4]{Number of totally performed energy evaluations divided by the number of totally performed transition state computations. The number of energy evaluations include the evaluations used for transition state searches, minimizations, softening and MD (if applicable).}
\footnotetext[5]{Number of runs in which lowest-barrier pathways could not be found before identifying $5\times 10^5$ distinct minima.}
\label{tab:perfres}
\end{table}
 \begin{figure}
 \includegraphics{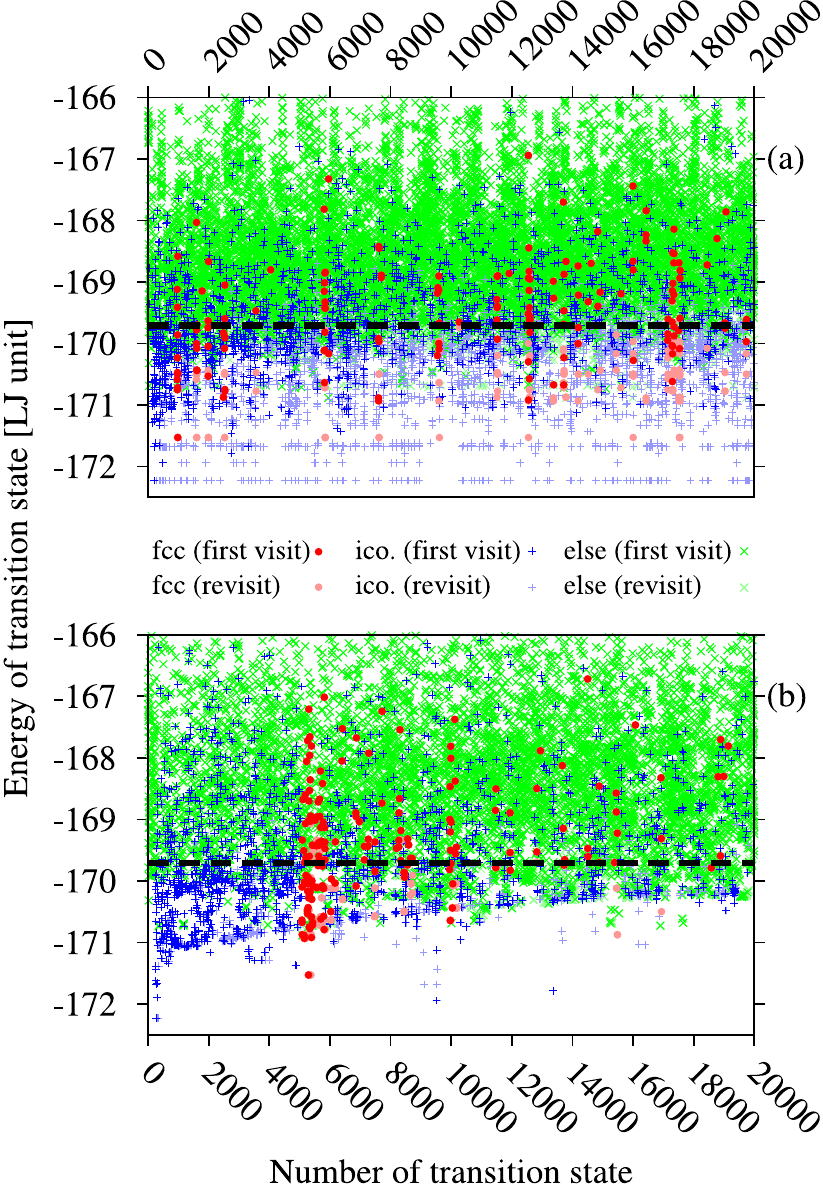}%
 \caption{\label{fig:history_lj38}Scatter plots showing all computed transition state energies in chronological order.
The shown data belongs to typical MHGPS and EFE runs for the $\mbox{LJ}_{38}$ two-funnel system.
Panel (a) shows MHGPS data, panel (B) shows EFE data.
Transition states belonging to the fcc funnel are represented by red $\bullet$ and transition states belonging to the icosahedral funnel are represented by blue +.
The green $\times$ represent all remaining transition states.
If a transition state is visited for the first time, the respective data point is dark-colored, otherwise it is light-colored.
The bold dashed line located at an energy of roughly $-169.709 \epsilon$ represents the highest barrier along the lowest-barrier pathway connecting the two energetically lowest minima of $\mbox{LJ}_{38}$.
An interpretation of this figure is given in the text of section \ref{sec:analysis}.}
 \end{figure}

In contrast to global minimum searches, a performance analysis of stationary point database generation algorithms is not straightforward since there is no obvious stopping criterion.
One possible stopping criterion can be defined by checking whether a putative lowest-barrier pathway between two minima has been found.
Because of the computational cost of Dijkstra's algorithm, this check is not feasible if it has to be performed between every pair of minima for a given system.
Therefore, a suitable test system should contain two outstanding and well defined minima for which pathways that connect them can be examined.
The global minimum of $\mbox{LJ}_{38}$ is located inside a small funnel containing fcc-like structures, the second-lowest minimum of $\mbox{LJ}_{38}$ is contained inside a comparatively large icosahedral funnel. Both funnels are separated by a high energy barrier.\cite{Doye1999,Doye1999-2}
Furthermore, the number of atoms in $\mbox{LJ}_{38}$ is small enough to perform a sufficient number of runs within a feasible amount of time.
Therefore, $\mbox{LJ}_{38}$ fulfills all requirements on being a suitable benchmark system.

Table \ref{tab:perfres} shows the results of a performance test based on 1000 independent runs for $\mbox{LJ}_{38}$.
Each run was started using a random non-fcc structure as input geometry and, depending on what happened earlier, was either stopped as soon as the putative lowest-barrier pathway between the global minimum and the second lowest local minimum of $\mbox{LJ}_{38}$  was identified, or if $5\times 10^5$ distinct local minima were found.
For all methods and all runs the same convergence criteria for the stationary points were used.

EFE needed roughly between a factor of 12 to 18 more transition state computations than the MHGPS method before encountering a lowest-barrier pathway of $\mbox{LJ}_{38}$.
Because the number of energy evaluations per transition state computation $n_{\mbox{E}}$ are similar for both methods, similar factors are obtained when measuring the computational cost in terms of energy evaluations.

For the EFE method we could observe a small number of runs that failed to find a lowest-barrier pathway at all.
Since the number of failure runs decreased with increasing number of followed mode directions these failures can be explained by the limited number of search directions available to the EFE method.
Assuming free boundary conditions, the EFE method can follow at maximum $6N-12$ directions per minimum for a $N$-atom system.
However, the number of transition states connected to a minimum can exceed the number of $6N-12$ directions by far.
For example it is known that the global minimum of $\mbox{LJ}_{13}$ is surrounded by 535 local minima which are connected to the global minimum by 911 transition states.\cite{Doye1999-2}
It is therefore possible to miss stationary points that potentially lie on the lowest-barrier pathway.
This general restriction of the EFE-method and similar deterministic mode-following methods has been mentioned before by Malek and Mousseau.~\cite{Malek2000}
By using random displacements away from the initial minimum, they showed that it is possible to avoid this problem in advanced mode-following techniques like the Activation Relaxation Technique.

The average number of distinct transition states $\left<n_{\mbox{ts,diff}}\right>$ divided by the average number of computed transition states $\left<n_{\mbox{ts}}\right>$ was between $66\%$ and $87\%$ larger for the MHGPS method than corresponding ratios of the EFE method.

The average CPU time required before MHGPS identified the lowest-barrier pathways between both lowest structures of $\mbox{LJ}_{38}$ was measured to be roughly 8 minutes (on a single core of an Intel Xeon E5-2665 CPU clocked at 2.40GHz).
This timing should be compared to the $10^5$ CPU hours that were required for the Sliding and Sampling computations reported in Ref.~\onlinecite{Miller2007}.
These timings differ by several orders of magnitude and therefore allow to give a rough idea on the performance differences between the different methods.
They are particularly noteworthy when noting that Ref.~\onlinecite{Miller2007} only presents pathways that are higher in energy than the known lowest-barrier pathway.\cite{Note1}
As well as MHGPS, the EFE method is also several orders of magnitudes faster than Sliding and Sampling.
On average, our implementation of EFE needed just under 3 CPU hours to find the lowest barrier path for $\mbox{LJ}_{38}$ ($n_{\mbox{ev}}=10$, average taken over successful runs).
As the CPU time depends very strongly on the computer hardware and the implementation of an algorithm, one should compare methods that do not exhibit such a distinct timing difference by using more suitable quantities like those given in Table~\ref{tab:perfres}.

Fig.~\ref{fig:history_lj38} shows the histories of all transition state energies of two typical MHGPS (panel (a)) and EFE (panel (b)) runs that were performed on the $\mbox{LJ}_{38}$ system.
Both runs were started at non-fcc structures and thus are residing inside the large icosahedral funnel during the first transition state computations.
Fig.~\ref{fig:history_lj38} illustrates the differing transition state sampling behavior of both methods.
In the very beginning the EFE method is able to sample low-energy transition states.
However, with an increasing number of totally sampled transition states, the energies of the lowest transition states that are being sampled also rises.
This means the EFE-method explores the energy landscape in a bottom-up fashion.
In conjunction with the limited number of search directions per minimum, this is a severe problem in particular for multi-funnel systems.
As can be seen from Fig.~\ref{fig:history_lj38}, in the beginning of the sampling procedure the bottom-up sampling forces a very detailed exploration of the icosahedral funnel. The EFE method is therefore not able to escape from the icosahedral to the fcc funnel until roughly 5000 transition state have been computed.
In very long runs, the same bottom-up sampling of the EFE method will lead at some point to the computation of transition states that almost entirely have energies above the highest barrier along the lowest-barrier pathway (energies above the bold, dashed and black line in Fig.~\ref{fig:history_lj38}).
If the lowest-barrier pathway could not be found up to that critical point, it is very unlikely that the EFE method will find it later.
In contrast to the EFE method, the MHGPS method escapes from the icosahedral to the fcc funnel very quickly and regularly switches back and forth between both funnels.
Because MHGPS does not strictly avoid previously visited low energy configurations, it does not suffer from the consequences of a strict bottom-up sampling.
MHGPS is always able to go down to previously explored low energy configurations, however the history based energy feedback takes care that well explored regions are left quickly.
Therefore, as illustrated by Fig.~\ref{fig:history_lj38}, MHGPS is able to sample transition states from the whole energy range at any stage of sampling.

We also looked at the 75-atom Lennard Jones system and found a similar behavior as for $\mbox{LJ}_{38}$. Starting at the second lowest minimum of $\mbox{LJ}_{75}$, which is contained in an icosahedral funnel, we performed MHGPS and EFE test runs which were stopped as soon as 275,000 transition states were computed.
Within this amount of computed transition states, our implementation of the EFE method showed not to be able to leave the icosahedral funnel, whereas the MHGPS method could switch between both $\mbox{LJ}_{75}$ funnels multiple times.

We also performed a short test run for the $\mbox{LJ}_{55}$ cluster which is a strong structure seeker.~\cite{Doye1999-2}
Despite its structure seeker character there exist two non-icosahedral minima which lie behind comparatively high barriers.~\cite{Wales1993,Doye1999-2,Wolf1998}
Each method's test run was started at the same arbitrarily chosen high energetic local minimum (-270.302962 $\epsilon$) and was stopped as soon as $30,000$ transition state computations were performed.
The overall appearance of the disconnectivity graph containing the lowest 700 minima generated from EFE-sampling is equivalent to the graph presented in Ref.~\onlinecite{Doye1999-2}, however in this test run our implementation of the EFE method could not identify the lower of the two non-icosahedral minima. 
The other of the two mentioned non-icosahedral minima could be found by the EFE method, however the barrier connecting it to the global minimum funnel was significantly larger than the barrier found in Ref.~\onlinecite{Doye1999-2}.
In contrast, the disconnectivity graph containing the lowest 700 minima generated from the MHGPS run contained all important features of the $\mbox{LJ}_{55}$ potential energy landscape, including both of the above mentioned non-icosahedral minima.
The barriers connecting the two non-icosahedral minima to the global minimum funnel were also reproduced in accordance with the barriers of the disconnectivity graph presented in Ref.~\onlinecite{Doye1999-2}.

\section{Application of MHGPS to $\mbox{LJ}_{75}$ and $\mbox{LJ}_{102}$}\label{sec:results}
 \begin{figure*}
 \includegraphics{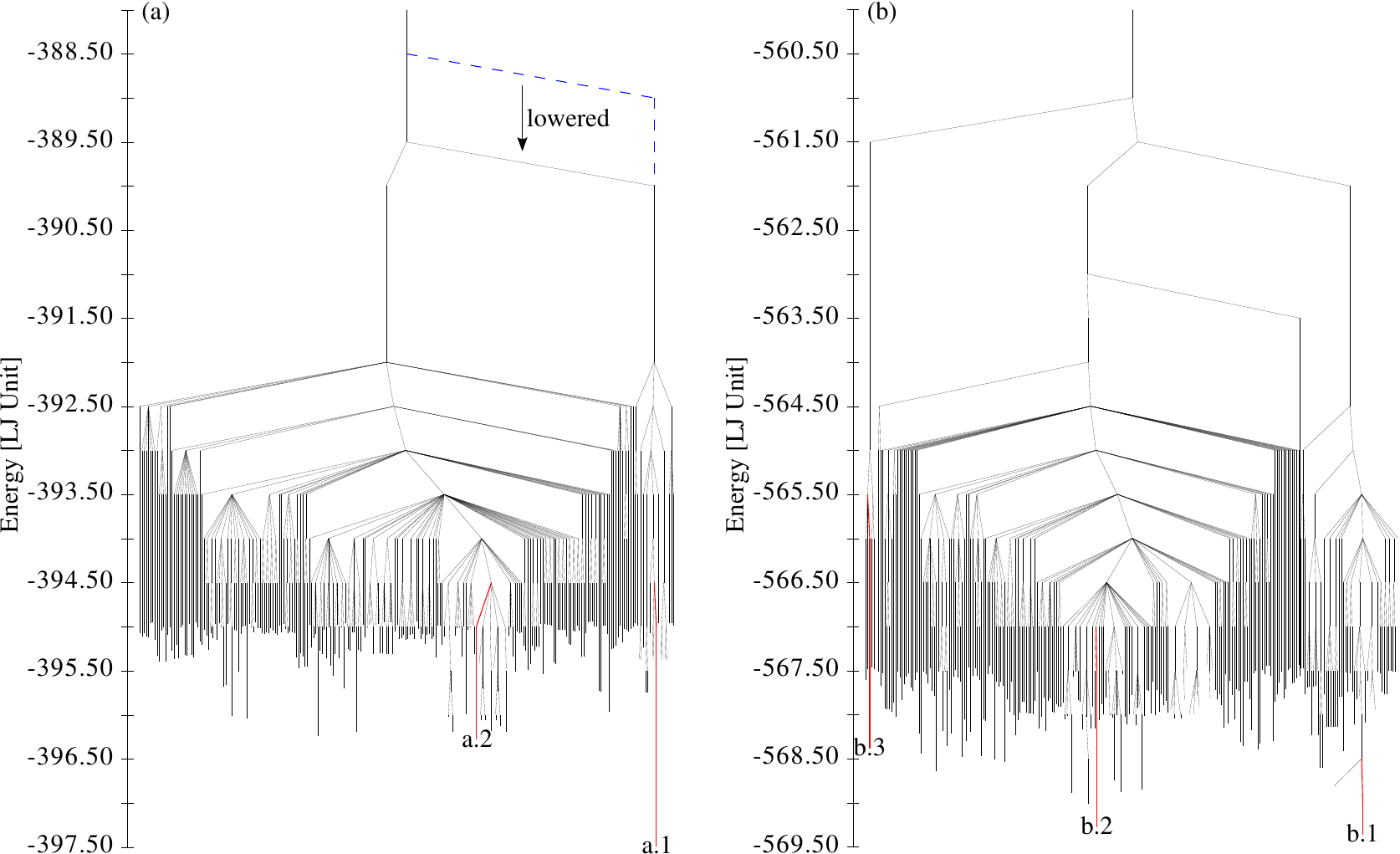}%
 \caption{\label{fig:trees} Disconnectivity graphs of $\mbox{LJ}_{75}$ [panel (a)] and $\mbox{LJ}_{102}$ [panel (b)].
Panel (a) shows the new putative lowest barrier between both funnels. The blue dashed line indicates the previously known lowest barrier connecting both funnels.~\cite{Doye1999-2}
Panel (b) shows a third, previously unknown, funnel with an energetically low bottom structure (minimum b.3) and a high barrier connecting it to the other two funnels.
Both graphs show the 250 lowest minima that were found for each system.
The bottom structures of each major funnel are labeled and highlighted using red color.}
 \end{figure*}
 \begin{figure}[!hb]
 \includegraphics{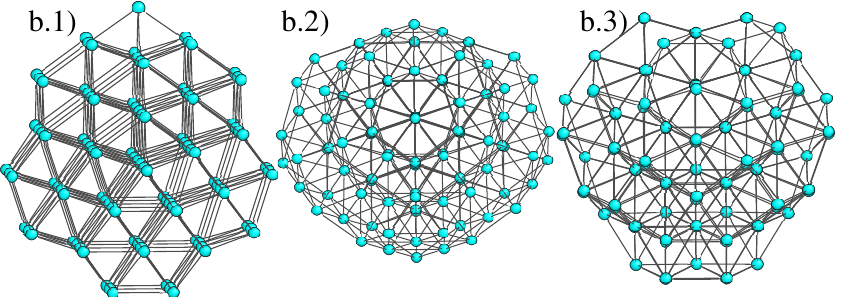}%
 \caption{\label{fig:lj102bottomstructures} Bottom structures of the three major funnels of $\mbox{LJ}_{102}$. The labeling of the illustrations corresponds to the labeling of panel (b) in Fig.~\ref{fig:trees}}
 \end{figure}
 \begin{figure*}
 \includegraphics{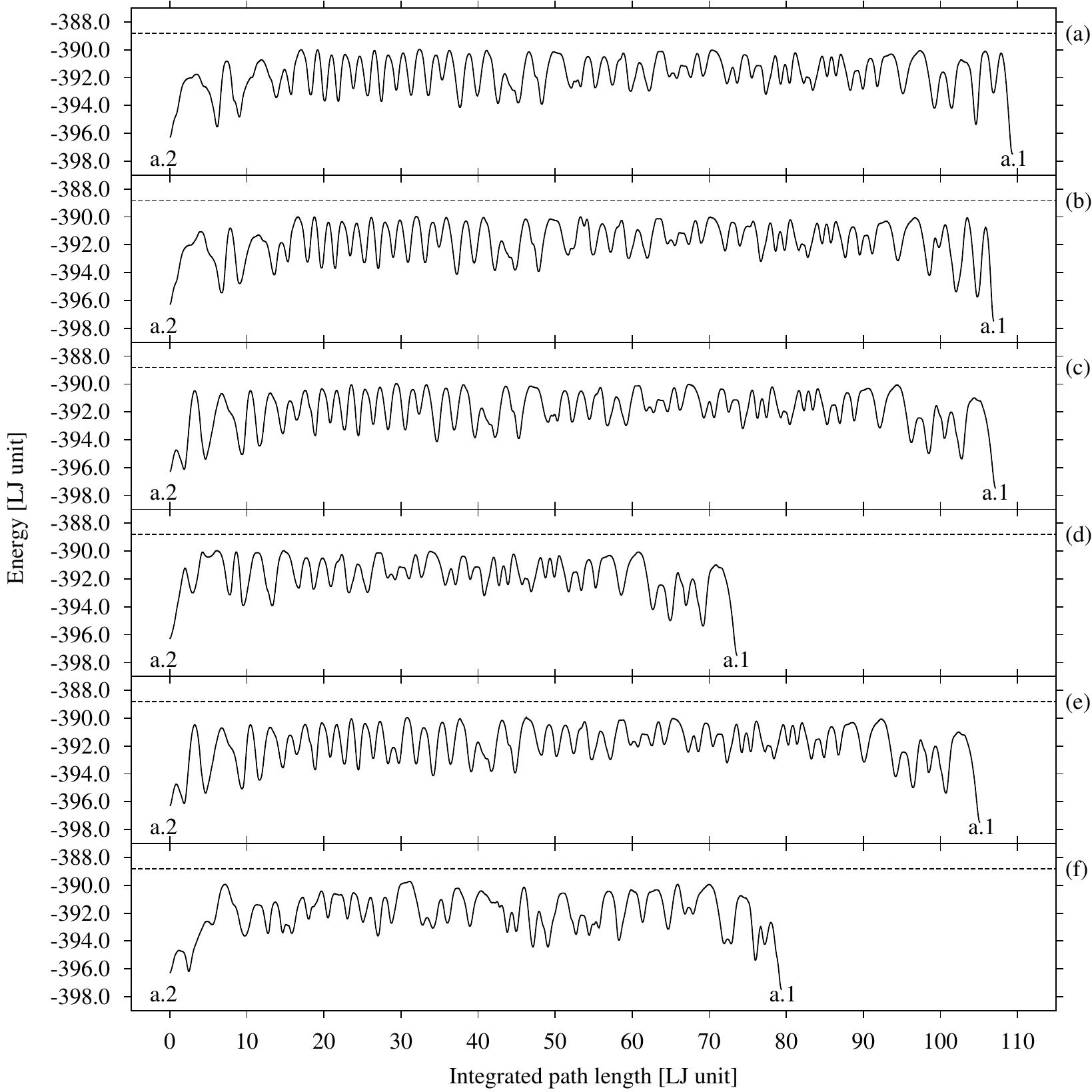}%
 \caption{\label{fig:pathslj75}Pathways found by MHGPS connecting the bottom-structures of both $\mbox{LJ}_{75}$ funnels (configurations a.2 and a.1 of Fig.~\ref{fig:trees}). The dashed horizontal lines indicate the highest energy along the previously known lowest-barrier pathway.~\cite{Doye1999-2}
 Panels (a), (b) and (c) show three alternative putative lowest-barrier pathways.
Panels (d), (e) and (f) show pathways that have been obtained by successively removing the highest energy transition state along the lowest-barrier pathway from the stationary point database [panels (d) and (e)] or from a preliminary test run [panel (f)].
They only have slightly higher barriers than the pathways of panels (a) to (c) and thus show that there exist a variety of pathways lying energetically between our best results and the previously presented~\cite{Doye1999-2} lowest-barrier pathways for $\mbox{LJ}_{75}$.}
 \end{figure*}
 \begin{figure*}
 \includegraphics{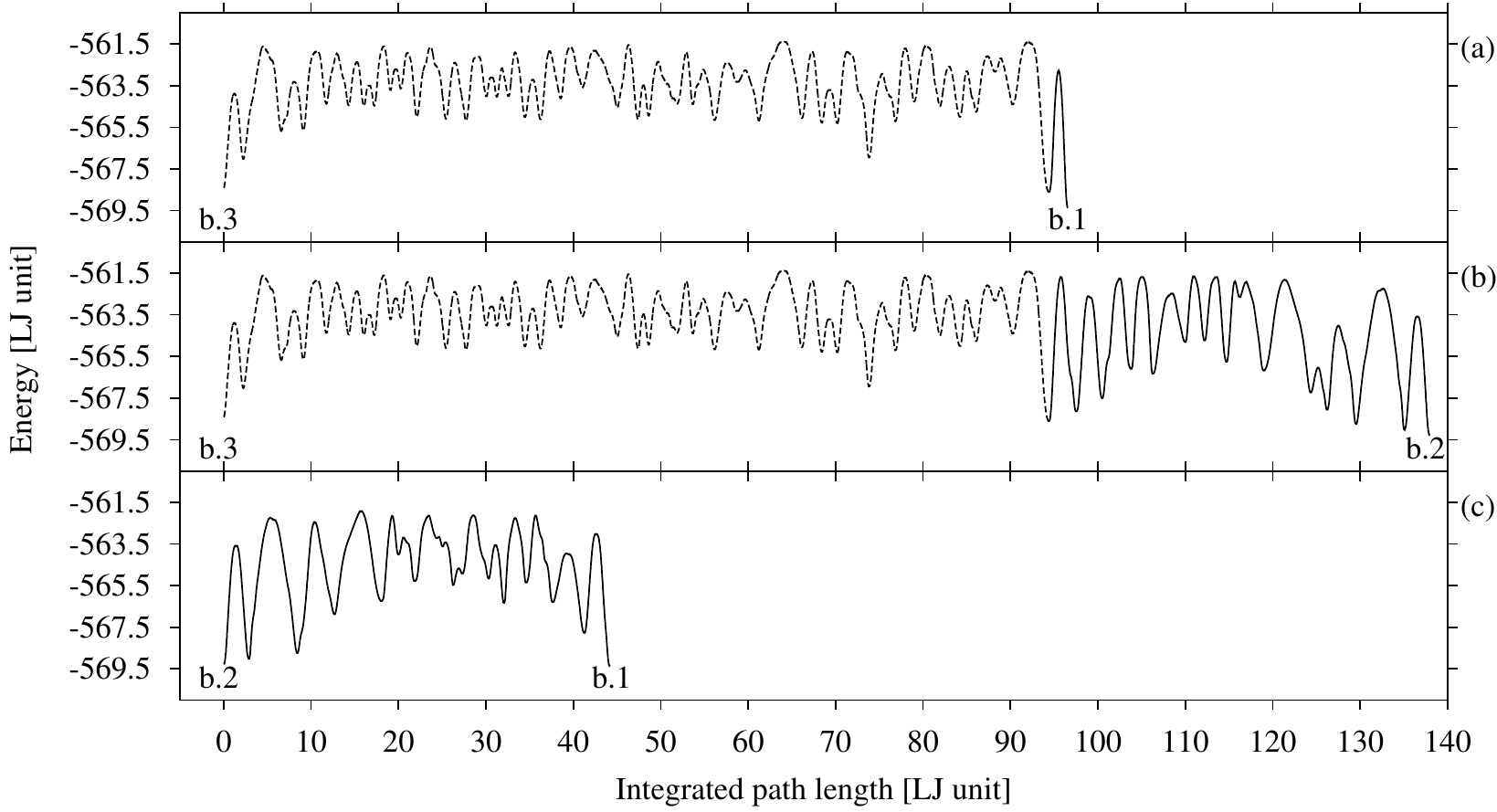}%
 \caption{\label{fig:pathslj102} Putative lowest-barrier pathways that were found by MHGPS for $\mbox{LJ}_{102}$.
Panel (a) shows a putative lowest-barrier pathway connecting the putative global minimum (configuration b.1 of Fig.~\ref{fig:trees}) to structure b.3 of Fig.~\ref{fig:trees}.
A lowest-barrier pathway connecting the second-lowest minimum of $\mbox{LJ}_{102}$ (configuration b.2 of Fig.~\ref{fig:trees}) and configuration b.3 of Fig.~\ref{fig:trees} is shown in panel (b).
The parts of the reaction pathways shown in panel (a) and (b) that coincide with each other are highlighted by using dashed lines.
Panel (c) shows a putative lowest-barrier pathway connecting the second-lowest configuration of $\mbox{LJ}_{102}$ (configuration b.2 of Fig.~\ref{fig:trees}) to the putative global minimum (configuration b.1 of Fig.~\ref{fig:trees}).
}
 \end{figure*}
Due to its advantages presented in section \ref{sec:analysis} we applied MHGPS to $\mbox{LJ}_{75}$ and $\mbox{LJ}_{102}$.
Concerning the task of sampling relevant stationary points, in particular $\mbox{LJ}_{75}$ is known to be a very difficult system.
This is explained by the frustration of its potential energy landscape and the large geometrical differences of both structures located at the bottoms of two major funnels.~\cite{Doye1999-2}

For each system we started 10 independent runs at the corresponding global minimum structures.
For every run different random seeds were used.
A run was stopped, as soon as $2\times 10^{6}$ distinct local minima were found.
For the analysis of the potential energy landscapes the stationary point databases resulting from all runs were merged into a single database for each system.
For $\mbox{LJ}_{75}$ this procedure resulted in a stationary point database containing roughly $12.0\times10^{6}$ distinct transition states connecting $7.0\times10^{6}$ distinct local minima.
In case of $\mbox{LJ}_{102}$  we obtained by this procedure a database containing roughly $10.9\times10^{6}$ distinct transition states which connect $7.5\times10^{6}$ distinct local minima.
The disconnectivity graphs of both system are shown in Fig.~\ref{fig:trees}.
Fig.~\ref{fig:pathslj75} and Fig.~\ref{fig:pathslj102} show plots of the energy along the reaction pathways in dependence of the integrated path length $S$ which is defined by the arc length of the steepest descent reaction path in the $3N$-dimensional coordinate space.\cite{Wales1994}
Numerically the integrated path length is computed by summing up all the lengths $\left|\Delta\mathbf{R}\right|$ of all steepest descent steps:
\begin{equation}
S=\sum_{\mbox{steps}} \left|\Delta\mathbf{R}\right|.
\end{equation}

\subsection{$\mbox{LJ}_{75}$}
As shown in panel (a) of Fig.~\ref{fig:trees} the highest barriers along the lowest-barrier pathways connecting the two major funnels of $\mbox{LJ}_{75}$ that were found by MHGPS are significantly lower in energy than those of the previously known lowest-barrier pathways.
Using Dijkstra's algorithm as outlined in section ~\ref{sec:transitionstates}, we could identify roughly 20,000 pathways all having the same highest-barrier energies of $7.51\epsilon$ and $6.30\epsilon$ and the same number of $51$ intermediate transition states.
Compared to this, the previously known lowest-barrier pathway has significantly higher highest-barrier energies of $8.69\epsilon$ and $7.48\epsilon$ and possesses $65$ intermediate transition states.~\cite{Doye1999-2}
In order to illustrate typical differences between alternative lowest-barrier pathways, the panels (a), (b) and (c) of Fig.~\ref{fig:pathslj75} explicitly show the steepest descent reaction paths of three lowest-barrier pathways.
In order to check whether there might exist further pathways which are energetically in-between the previously known lowest-barrier pathway and the putative lowest-barrier pathways found by MHGPS, we successively removed the highest energy transition state along the lowest-barrier pathway from the stationary point database and applied Dijkstra's algorithm.
Pathways resulting from this removal are shown in panels (d) and (e) of Fig.~\ref{fig:pathslj75}.
For the pathway shown in panel (d) the barriers are $7.52\epsilon$ and $6.31\epsilon$, for the pathway of panel (e) the barriers are $7.54\epsilon$ and $6.33\epsilon$.
They are only slightly higher in energy than the highest barriers along the putative lowest-barrier pathway.
This suggests that there exists a whole range of pathways that are energetically between the putative lowest pathways presented in this study and the previously known lowest pathway.
This conjecture seems to be reinforced by the pathway shown in panel (f) of Fig.~\ref{fig:pathslj75}.
This pathway was found in a preliminary single-run test in which only roughly $6\times 10^{5}$ distinct local minima and roughly $9\times10^{5}$ distinct transition states were sampled.
The highest barriers along this pathway are $7.78 \epsilon$ and $6.57\epsilon$.

\subsection{$\mbox{LJ}_{102}$}
As shown in panel (b) of Fig.~\ref{fig:trees} MHGPS could find a previously unknown funnel for $\mbox{LJ}_{102}$.~\cite{Doye2006}
An illustration of the bottom structure of this funnel is given in Fig.~\ref{fig:lj102bottomstructures}.
The new bottom structure possesses icosahedral elements and its surface is dominated by buckled hexagonal patches. Its has an energy of $-568.388773\epsilon$.

Lowest-barrier pathways connecting the new structure to the global minimum and to the second lowest minimum are shown in panels (a) and (b) of Fig.~\ref{fig:pathslj102}.
The lowest-barrier pathways connecting this new structure and the global minimum contain $40$ intermediate transition states and the highest barriers are $7.97\epsilon$ and $7.89\epsilon$.
The highest barriers of the lowest-barrier pathways that connect the second lowest minimum to the bottom of the new funnel are $7.97\epsilon$ and $7.00\epsilon$. These pathways contain $53$ intermediate transition states.

Furthermore, MHGPS could confirm the energy of the highest barrier along the putative lowest-barrier pathway connecting the global minimum to the second lowest minimum.~\cite{Doye2006}
However, both in terms of the number of intermediate transition states and in terms of the integrated path length, the pathway found by MHGPS is significantly shorter than the previously known pathway.
It contains only 16 intermediate transition states compared to 30 transition states contained in the pathway published earlier~\cite{Doye2006}. The integrated path length is roughly 11$\sigma$ shorter (difference of paths length was estimated using the plot of Ref.~\onlinecite{Doye2006}).

\section{Conclusion}\label{summary}

MH is a practical guide for the search of low-barrier reaction pathways, because it uses short MD moves for the exploration of potential energy surfaces and an energy feedback that satisfies the explosion condition~\cite{Goedecker2004,Goedecker2011}.
As a consequence of the short MD moves, consecutive minima along the MH trajectory are structurally not too different from each other and thus are well suited as input structures for methods that can find transition states between two given input geometries. Furthermore, energy conservation assures that the maximum barrier energy between two consecutive minima is bounded from above.
The explosion condition assures that the MH guide does not get stuck in deep funnels.
As a consequence, MHGPS must perform computationally expensive transition states computations only between minima that are particularly promising for the purpose of finding energetically low barriers and between minima that are promising for the exploration of the potential energy landscape.
MHGPS needs no human intuition and its MH based exploration of the potential energy surface is completely unbiased. It therefore does not not fail to explore unforeseen and unexpected features of potential energy landscapes.
In comparison to the EFE mode-following approach, MHGPS detects a significantly larger number of distinct transition states when performing the same number of transition state computations.
MHGPS reduces the cost of sampling stationary points and their connectivity information by over one order of magnitude compared to the EFE mode-following approach.
In contrast to other methods, MHGPS could successfully find the lowest-barrier pathways of $\mbox{LJ}_{38}$ in all tests.
The efficiency of our new method is also confirmed by new results that were found for $\mbox{LJ}_{75}$ and $\mbox{LJ}_{102}$, systems that have been thoroughly examined for more than a decade.

\begin{acknowledgments}
We thank the Indo Swiss Joint Research Programme (ISJRP) for financial support. Parts of the development of Bar-Saddle were supported by a grant from the Swiss National Supercomputing Centre (CSCS) under project ID s142.
\end{acknowledgments}

\appendix
\section{Stability of the modes}
\label{app:Stability_of_the_modes}

The curvature along an arbitrary vector $\mathbf{x}$, evaluated at the position $\mathbf{x}_0,$ is defined as
\begin{equation}
 C_{\mathbf{x}_0}(\mathbf{x})=\frac{\mathbf{x}^T\mathbf{H}_{\mathbf{x}_0}\mathbf{x}}{\mathbf{x}^T\mathbf{x}},
 \label{eq:definition_curvature}
\end{equation}
where $\mathbf{H}_{\mathbf{x}_0}$ is the Hessian at $\mathbf{x}_0$. If $\mathbf{x}$ was an eigenvector $\mathbf{v}_i$, this would give the corresponding eigenvalue $\lambda_i$.
Furthermore, calculating the gradient with respect to $\mathbf{x}$ under the constraint of normalization gives
\begin{equation}
\frac{1}{2}\frac{\mathrm{d}}{\mathrm{d}\mathbf{x}} \frac{\mathbf{x}^T\mathbf{H}_{\mathbf{x}_0}\mathbf{x}}{\mathbf{x}^T\mathbf{x}}\bigg|_{\mathbf{x}^T\mathbf{x}=1} = \mathbf{H}_{\mathbf{x}_0}\mathbf{x} - \left(\mathbf{x}^T\mathbf{H}_{\mathbf{x}_0}\mathbf{x}\right) \mathbf{x}.
\end{equation}
This expression vanishes in case $\mathbf{x}$ is an eigenvector, showing that the eigenmodes are stationary points of $C_{\mathbf{x}_0}(\mathbf{x})$.

The next point is to show that among all these stationary directions, only the lowest mode is actually a minimum and thus stable, meaning that rotating a slightly misaligned dimer according to its torsional force will lead back to this mode.
Since the eigenvectors form a complete set, any vector can be written as a linear combination of them, i.e.\ $\mathbf{x}=\sum_i c_i\mathbf{v}_i$,
with the normalization condition $\sum_i c_i^2=1$. Plugging this into Eq.~\eqref{eq:definition_curvature} and using the orthonormality of the eigenvectors gives
\begin{equation}
 C_{\mathbf{x}_0}(\mathbf{x}) = \sum_i c_i^2\lambda_i = c_l^2\lambda_l + c_m^2\lambda_m + c_n^2\lambda_n + \sum_{i \notin \{l,m,n\}}c_i^2\lambda_i.
 \label{eq:curvature_in_eigenbasis}
\end{equation}
There are three cases to consider:
\begin{itemize}
 \item $m$ corresponds to the lowest eigenvalue: Eq.\eqref{eq:curvature_in_eigenbasis} is minimal for the set $\{c_l=0,c_m=1,c_n=0,c_i=0\}$,
  proving that the lowest mode corresponds to a minimum.
 \item $m$ corresponds to the highest eigenvalue: Eq.\eqref{eq:curvature_in_eigenbasis} is maximal for the set $\{c_l=0,c_m=1,c_n=0,c_i=0\}$,
  proving that the highest mode corresponds to a maximum.
 \item $m$ corresponds neither to the lowest nor the highest eigenvalue: assuming $\lambda_l<\lambda_m<\lambda_n$, then choosing $\{c_l=1,c_m=0,c_n=0,c_i=0\}$ results in $C<\lambda_m$,
  whereas choosing $\{c_l=0,c_m=0,c_n=1,c_i=0\}$ results in $C>\lambda_m$.
  Together this shows that all these modes are saddle points.
\end{itemize}

\providecommand{\noopsort}[1]{}\providecommand{\singleletter}[1]{#1}%

\end{document}